\documentclass[reqno,11pt]{amsart}

\usepackage{amsmath,amssymb,amsthm}
\usepackage[usenames,dvipsnames]{color}
\usepackage[bookmarks,colorlinks,breaklinks]{hyperref}  
\usepackage{doi,url}
\usepackage{enumerate}
\usepackage{bbm}
\usepackage{mathabx}
\usepackage{appendix}

\definecolor{dullmagenta}{rgb}{0.4,0,0.4}   
\definecolor{darkblue}{rgb}{0,0,0.4}
\hypersetup{linkcolor=red,citecolor=magenta,filecolor=dullmagenta,urlcolor=darkblue}

\numberwithin{equation}{section}

\newtheoremstyle{ttheorem}%
       {1.3ex\@plus1ex}                
       {2.5ex\@plus1ex\@minus.5ex}      
       {\itshape}           
       {0pt}                   
       {\bfseries}          
       {.}                  
       {.5em}               
       {}                

\newtheoremstyle{ddefinition}%
       {1.3ex\@plus1ex}                
       {2.5ex\@plus1ex\@minus.5ex}      
       {}           
       {0pt}                   
       {\bfseries}           
       {.}                  
       {.5em}               
       {}                

\newtheoremstyle{rremark}%
       {1.3ex\@plus1ex}                
       {2.5ex\@plus1ex\@minus.5ex}      
       {\itshape}        
       {0pt}                   
       {\bfseries}           
       {.}                  
       {.5em}               
       {}                   

\theoremstyle{ttheorem}
\newtheorem{theorem}{Theorem}[section]

\newtheorem{rem}[theorem]{Remark}

\theoremstyle{ddefinition}

    {\end{nummer}\end{myremarks}}

\newcounter{numcount}
\newcommand{\labelnummer}{\mbox{\normalfont (\roman{numcount})}}%

\makeatletter

\newenvironment{nummer}%
  {\let\curlabelspeicher\@currentlabel%
    \begin{list}{\labelnummer}%
      {\usecounter{numcount}\leftmargin0pt%
        \topsep0.5ex\partopsep2ex\parsep0pt\itemsep0ex\@plus1\p@%
        \labelwidth2.5em\itemindent3.5em\labelsep1em%
      }%
    \let\saveitem\item%
    \def\item{\saveitem%
      \def\@currentlabel{\curlabelspeicher$\,$\labelnummer}}%
    \let\savelabel\label%
    \def\label##1{\savelabel{##1}%
      \@bsphack%
        \ifmmode\else%
          \protected@write\@auxout{}%
          {\string\newlabel{##1item}{{\labelnummer}{\thepage}}}%
        \fi%
      \@esphack%
    }%
  }{\end{list}}%

\def\itemref#1{\expandafter\@setref\csname r@#1item\endcsname%
  \@firstoftwo{#1}}%

\def\section{\@startsection{section}{1}%
  \z@{1.3\linespacing\@plus\linespacing}{.5\linespacing}%
  {\normalfont\scshape\centering}}

\DeclareMathOperator{\tr}{tr}

\renewcommand\L{\mathrm{L}}

\newcommand\RR{\mathbb{R}}
\newcommand\NN{\mathbb{N}}
\newcommand\ZZ{\mathbb{Z}}

\newcommand\DD{\mathbb{D}}

\renewcommand\P{\mathbb P}
\newcommand\E{\mathbb E}

\usepackage[active]{srcltx}
\theoremstyle{ttheorem}
 \newtheorem{thm}{Theorem}[section]

 \theoremstyle{definition}
 \newtheorem{defn}[thm]{Definition}

 \theoremstyle{remark}

 \numberwithin{equation}{section}

\newcommand{\be}{\begin{equation}}
\newcommand{\ee}{\end{equation}}
\newcommand{\benon}{\begin{equation*}}
\newcommand{\eenon}{\end{equation*}}
\newcommand{\ba}{\begin{array}}
\newcommand{\ea}{\end{array}}
\newcommand{\bal}{\begin{align}}
\newcommand{\eal}{\end{align}}
\newcommand{\bea}{\begin{eqnarray}}
\newcommand{\eea}{\end{eqnarray}}
\newcommand{\bee}{\begin{eqnarray*}}
\newcommand{\eee}{\end{eqnarray*}}

\newcommand{\Rd}{\mathbb R^d}
\newcommand{\Zd}{\mathbb Z^d}
\newcommand{\norm}[1]{\Vert #1 \Vert}
\newcommand{\abs}[1]{\left| #1 \right|}

\newcommand{\Lp}[1]{\textrm{L}^2(#1)}
\renewcommand{\L}{\Lambda}

\newcommand{\hm}[1]{\leavevmode{\marginpar{\tiny%
$\hbox to 0mm{\hspace*{-0.5mm}$\leftarrow$\hss}%
\vcenter{\vrule depth 0.1mm height 0.1mm width \the\marginparwidth}%
\hbox to
0mm{\hss$\rightarrow$\hspace*{-0.5mm}}$\\\relax\raggedright #1}}}

\newcommand{\vertiii}[1]{{\left\vert\kern-0.25ex\left\vert\kern-0.25ex\left\vert #1
    \right\vert\kern-0.25ex\right\vert\kern-0.25ex\right\vert}}

\begin{document}

\title[Random Schr\"odinger Operators in aperiodic media]{Random Schr\"odinger Operators and Anderson localization in aperiodic media}

 \author[Constanza Rojas-Molina]{C. Rojas-Molina}
 \address[]{D\'epartement de Math\'ematiques, Universit\'e de Cergy-Pontoise, CNRS UMR 8088, 2 avenue Adolphe Chauvin, 95302 Cergy–Pontoise, France}
\email{crojasmo@u-cergy.fr}
\maketitle

\begin{abstract}
In this note we review some results on localization and related properties for random Schr\"odinger operators arising in aperiodic media. These include the Anderson model associated to disordered quasycrystals and also the so-called Delone operators, operators associated to deterministic aperiodic structures.
\end{abstract}
\section{Introduction}

Disordered quantum systems has been an intensive field of research since the ground-breaking work of P. W. Anderson in the late 1950s \cite{And58}. This was the first description of the absence of electron propagation in crystals with impurities, a phenomenon known as Anderson localization. Since the late 1970s, mathematical-physicists have devoted efforts to establish a rigourous framework to describe localization, initiating the theory of random Schr\"odinger operators. By now, the effect of disorder on the wave propagation in materials is to a large extent well understood, both in the classical and quantum settings, see e.g. \cite{KK04} and \cite{GKuniv}, where localization appears whenever randomness is present. In the quest for understanding the robustness of localization, it is natural to consider models that differ from the disordered crystals studied in the literature, the so-called Anderson model.

In this note, we are interested in random quantum systems with underlying aperiodic structures. Aperiodic media is characterized by a lack of translation invariance in its atomic structure. Quasicrystals are a particular example of aperiodic media that lacks periodicity, but that exhibit instead some long range order, which justifies their name. Such materials have fascinated the mathematics and physics community since their discovery in 1982 by D. Shechtman. In \cite{Sch84} the authors showed for the first time that aperiodic structures, some of which thought to exist only theoretically, could be observed in physical experiments. This breakthrough gave an additional motivation to the study of aperiodic order, a field that was already present earlier in the literature, with for example, the notion of almost-periodic functions, or Penrose tilings. For references and more details on the discovery and state of the art of aperiodic order, see \cite{BaGr}.



We now introduce some notation.
Let $d\geq1$ be the space dimension and $\Lambda_{L}(x) := \raisebox{-.3ex}{\LARGE$\times$}_{j=1}^{d} (x_{j} - L/2, x_{j}+ L/2)$ be the open cube in $\Rd$ of side-length $L>0$ centered at $x=(x_{1},\ldots,x_{d})\in\RR^{d}$.
\begin{defn} \label{defdelone}
A point set $D$ of $\Rd$ is called an $(r,R)$-\emph{Delone set}
if ~(i)~~ it is \emph{uniformly discrete}, i.e.\ there exist a real $r>0$ such that $\big|D\cap\L_r(x)\big|\leq 1$  for every $x\in\Rd$, and ~(ii)~~ it is \emph{relatively dense}, i.e.\ there exists a real $R\geq r$ such that $\big|D\cap\L_R(x)\big|\geq 1$ for every $x\in\Rd$. Here, $|\cdot|$ stands for the cardinality of a set.
\end{defn}
Particular examples of Delone sets are the lattice $\mathbb{Z}^{d}$, the vertices of a Penrose tiling or the random point set obtained from removing every other point of $\ZZ$ by a Bernoulli percolation process.

Next, we consider we consider a Schr\"odinger operator with
a potential that encodes the interaction between the electron and the aperiodic environment. Throughout the text we will focus on the continuous case $\Lp{\Rd}$. For the discrete case, we refer the reader to \cite[Section 2.3.1]{RMch}.

\begin{defn}\label{defrandop}
Given an $(r,R)$-Delone set $D$ in $\Rd$, we define the Delone-Anderson model as the random Schr\"odinger operator
\be\label{hdom} H_{D^\omega}:= H_0+V_{D^\omega} \ee
with dense domain in the Hilbert space $\Lp{\Rd}$ verifying:
\begin{itemize}
\item[i)]
	The background operator is either the negative Laplacian $H_0 :=-\Delta$ or $H_0:=-\Delta+V_0$ with $V_0$ a non-negative, bounded, measurable potential.
\item[ii)] The random potential is given by
	\be \label{ranpot1}
		\Rd \ni x \mapsto V_{D^\omega}(x) := \sum_{\gamma \in D}\omega_\gamma u(x-\gamma)
	\ee
	with $u$ a non-negative, continuous \emph{single-site potential}, with compact support such that the supports of $u(\cdot-\gamma)$ do not overlap. The
	(canonically realized) random variables $\omega:=(\omega_\gamma)_{\gamma\in D}$ are
	 i.i.d., distributed according to the probability measure $\P_0$ with compact support $\mathbb A\subset \RR$. We assume that $0\in \mathbb A$, and denote the product probability space
	$\Omega_{D}:=\bigtimes_{D}\mathbb A$, with probability measure $\P_D$.  We denote by $\E_{D}$ the expectation with respect to $\P_D$.
\end{itemize}
\end{defn}
The particular case $D=\Zd$ defines the usual alloy-type Anderson model associated to disordered crystals, while the case of arbitrary Delone set $D$ models disordered aperiodic media. If we assume certain conditions on $D$ to reflect long range order, then  \eqref{hdom} is the operator associated to a disordered quasicrystal.

In $d=2$, we also allow for $H_{D^\omega}$ to be the random Landau Hamiltonian given by
\be\label{magnop} H_{D^\omega}= (-i\nabla-{A})^2 + V_{D^\omega}, \ee
 with constant magnetic field $B>0$, vector potential $A:=\frac{B}{2}(x_2,-x_1)$ in the symmetric gauge, and $V_{D^\omega}$ as in \eqref{ranpot1} (for details on the magnetic case and definitions, see \cite{RM12}).

In the rest of this note we will review some particular results on disordered quasicrystals. On one hand, we will see that the lack of translation invariance has strong consequences in the ergodic properties of the model, but that quasicrystals and crystals share the same features due to the kind of long-range order they exhibit. This might not be necessarily the case for more general aperiodic media. On the other hand, we will see that in arbitrary aperiodic media, disorder  still produces localization, and that actually localization is insensitive to the underlying configuration of atoms, as long as it homogeneous, in the sense of a Delone set configuration. This implies in particular localization for disordered quasicrystals. Finally, we will see how the results on Delone-Anderson models can be applied to study the existence of Delone sets for which the associated Schr\"odinger operators
\be H_D=-\Delta + V_D,\quad \mbox{with}\,\, V_D=\sum_{\gamma \in D}u(x-\gamma), \ee
exhibit localization at the bottom of their spectrum.
 The theory of (disordered) aperiodic media is a rich field that combines approaches coming from diverse disciplines, like dynamical system, spectral theory, or operator algebras. In this note we focus on an approach that exploits the many developments that the theory of random Schr\"odinger operators has seen in recent years.

\vspace{-0.3cm}
\section{Ergodic properties of random Delone operators}

Ergodicity is a property of random Schr\"odinger operators with far-reaching consequences in describing almost-sure properties. In the case of the Anderson model $D=\Zd$, ergodicity is a consequence of the underlying periodic lattice structure of the potential and the iid nature of the random coupling constants. This implies that there exist measure-preserving ergodic transformations $\{\tau_{a}\}_{a\in\Zd}$ on $\Omega_{\Zd}$ and a family of unitary translation operators $\{U_a\}_{a\in\Zd}$ acting on $\mathrm{L}^2(\Rd)$ such that the following covariance relation holds
\be\label{ergo}   H_{(\Zd)^{\tau_{a}(\omega)}}= U_a H_{(\Zd)^{\omega}} U_a^*\ee
for every $a\in\Zd$. This property implies, in particular, that the spectrum $\sigma(H_{(\Zd)^{\omega}})$ of the elements of the family $(H_{(\Zd)^{\omega}})_{\omega\in\Omega}$ is deterministic, i.e., that there exists as set $\Sigma\subset \RR$ such that $\Sigma=\sigma(H_{(\Zd)^{\omega}})$ for $\P_{\Zd}$-a.e. $\omega\in\Omega$, see e.g. \cite{GMP77,Pas80}. This makes the statements on almost-sure localization meaningful, in the sense that the results are obtained for almost every operator in the family in a region of energies that is inside the spectrum, almost-surely.
Ergodicity also allows for the application of ergodic theorems to show the existence of the integrated density of states (IDS), a function that is of especial interest in the proofs of localization.

In the aperiodic case, the lack of translation invariance of $D$ yields a break of ergodicity, loosing \eqref{ergo}.
Despite this, we can still study the ergodic properties of Delone-Anderson operators, relying on some geometric properties of the underlying Delone set instead of its periodicity.
The long range-order in Delone sets can be expressed by the repetition of patterns in the set. A {\em pattern} is any set of the form $K\cap D$, with $K\subset $ a compact set. The following notion is a sufficient condition for an aperiodic set $D$ to have the same ergodic properties as the periodic case $D=\Zd$.
\begin{defn}(spUPF)\label{def-upf}
The Delone set $D$ has \emph{uniform pattern frequency} if for any pattern $Q\subset D$ the quotient
\be
	\label{upf}
	 \frac{1}{L^d}\big| \left\{ \tilde Q\subset D\,:\, \exists y\in (x+\L_L) \,\,
	\mbox{ such that  }\,y+\tilde Q=Q \right\}\big|
\ee
converges uniformly in $x\in \Rd$ as $L \to\infty$.  Moreover, we say that $D$ has  \emph{strictly positive} uniform pattern frequencies (spUPF) if this limit is strictly positive.
\end{defn}
With the spUPF property, the condition $0\in\mathbb A$ on the probability distribution of the random variables in our model is enough to imply that
\be\label{eq:freespec} \sigma(H_0)\subset \sigma(H_{D^\omega}) \quad \P\mbox{-a.s.}\ee
This can be seen by using a Weyl-sequence argument, but replaces translation invariance by pattern repetition, see e.g. \cite[Theorem 2.10]{RMch}.
While \eqref{eq:freespec} is enough to show the existence of deterministic spectrum in the case $H_0=-\Delta$ in Definition \eqref{defrandop}, this is not enough to obtain a meaningful statement in the magnetic case \eqref{magnop}. This motivates the search for other ways to show the existence of a deterministic spectrum, which was done in \cite{GMRM} via the integrated density of states.

Note that if for $x\in \Rd$ we define conveniently translated sets $x+D=\{\gamma+x,\gamma\in D\}$ and for $D^{\omega}=\{(\gamma,\omega_\gamma),\gamma\in D,\omega_\gamma\in A\}$ in such a way that $x+D^{\omega}=(x+D)^{\tau_{x}(\omega)}$, we see that
\be\label{deloneergo}   H_{(x+D)^{\tau_{x}(\omega)}}=  U_x H_{D^{\omega}} U_x^*,\ee
with which we retrieve the covariance condition from \eqref{ergo}. This observation can be made rigorous, and is a consequence of ergodicity in the framework of operators associated to randomly colored Delone sets \cite{MR}. There is a vast literature on Delone sets from the perspective of the set of all possible translations of $D$, which generates the so-called Delone dynamical systems. It is the dynamical aspects of Delone sets that justify the requirement for the spUPF in our setting. For more details, see e.g. \cite{KLS03a,LS03, BaGr} and references therein.


We can now introduce the finite-volume IDS, defined for a Delone set $D$  as
\be \RR\ni E\mapsto \nu_{ D^\omega,x,L}(E)=\frac{1}{L^{d}} \,
	\tr \left( \chi_{]-\infty,E]}(H_{D^{\omega}})\chi_{\L_{L}(x)}\right),  \ee
where $ \chi_{]-\infty,E]}(H)$ denotes the spectral projection of an operator $H$ on the interval $]-\infty,E]$, and $\chi_{\L_{L}(x)}$ is the characteristic function of the cube $\L_{L}(x)$.

The following is a particular case of the more general \cite[Corollary 2.8]{GMRM}. This special case is valid for operators associated to quasicrystals with impurities.
\begin{thm}[Existence of the IDS and a.s. spectrum] Let $D$ be a Delone set satisfying the spUPF property, and let either $H_{D^{\omega}}=-\Delta+V_{D^{\omega}}$ be given in \eqref{hdom}, or $H_{D^{\omega}}$ given in \eqref{magnop}. Then there exists a right-continuous non-decreasing function $\nu_{D}$, called the integrated density of states, such that for every $x\in\Rd$  we have \vspace{-0.3cm}
\be	\lim_{L\to\infty} \nu_{{{D}^{\omega},x,L}}(E) = \nu_{D}(E) \ee
for every continuity point of $\nu_{D}$.
\end{thm}
This theorem yields in particular that the spectrum $\sigma(H_{D^{\omega}})$ is deterministic, see \cite{GMRM}. The idea of the proof is to enlarge the space of parameters $\Omega_D$ to the randomly colored convex hull of $D$, that is defined by the closure $\hat X_D$ of all translates of ${D}^{\omega}=\{(\gamma,\omega_\gamma),\gamma\in D,\omega_\gamma\in \mathbb A\}$. Then we study the family of operators $(H_{P^{\omega}})_{P^{\omega}\in \hat X_D}$. In this setting we can use appropriate ergodic theorems obtained in \cite{MR} to show the existence of the limit of the finite-volume IDS for almost every element $P^{\omega}\in \hat X_D$ with respect to some probability measure $\hat\mu$ defined on $\hat X_D$. The strictly positive uniform pattern frequency property ensures que unique ergodicity of $X_D$, which allows to remove the $\hat\mu$-exceptional set in the results. The statements then holds for all $P\in \hat X_D$ and $\omega\in \Omega_P$ including the original Delone set $D$ we are interested in. See \cite[Section 2]{GMRM} for definitions and details.

Once the existence of the IDS has been shown, the next step is to study its behavior near the spectral gaps. For the case of $H_0=-\Delta$, we state a particular case of \cite[Theorem 3.1]{GMRM},

\begin{thm}[Lifshitz tails \label{LT}]
Let $D$ be a Delone set satisfying the spUPF and let $H_{D^{\omega}}=-\Delta+V_{D^{\omega}}$ be given in \eqref{hdom}  Assume, moreover, that the probability measure $\P_0$ is absolutely continuous with a regular enough probability density. Then, the IDS exhibits a Lifshitz tail at the bottom of the spectrum, i.e.,
\be \lim_{E\searrow0}\frac{\ln \abs{\ln \nu_D(E)}}{\ln E}=-\frac{d}{2}. \ee
\end{thm}
The proof relies on an averaged version of the Dirichlet-Neumann bracketing over the dynamical system generated by the set $D$.

\section{Localization for random Delone operators}
\begin{defn}
We say that the operator $H_{D^{\omega}}$ exhibits dynamical localization in an interval $I$ in its almost-sure spectrum if, for all $p\geq 0$ and any initially localized function $\varphi_0\in\Lp{\Rd} $, the following holds
\be\label{eq:at}\mathbb E_D\left(\sup_{t\in\mathbb R}\norm{\abs{X}^{p/2} e^{-itH_{D^{\omega}}}\chi_{I}(H_{D^{\omega}})\varphi_0}_2^2\right)<\infty,  \ee
where $\abs{X}$ is the multiplication operator defined by $\abs{X}\varphi(x)=\sqrt{1+\norm{x}^2}\varphi(x)$, and $\norm{\cdot}_2$ denotes the Hilbert-Schmidt norm.
\end{defn}
Dynamical localization (Anderson localization according to physicists), implies in particular pure point spectrum and exponentially decaying eigenfunctions (Anderson localization according to mathematicians). The unfortunate choice of terms is due to historical reasons, which might lead to some confusion when the mathematics and physics literature is compared. In this note we have intentionally avoided adjectives to the term {\em localization}.

Localization for Delone-Anderson operators in the continuous setting has been shown in the literature using both the Fractional Moment Method (FMM) \cite{BdMNSS06} and the Multiscale Analysis (MSA), see e.g. \cite{RM12, GMRM, MRM}. In \cite{RM12} it was shown that the MSA  is insensitive to perturbations of the underlying array of atoms, as long as it is homogeneous in the sense of being a Delone set. The advantage of the MSA is that allows for singular probability measures, like Bernoulli, a fact that was used in a crucial way to study purely (deterministic) aperiodic media, as we will see in the next section.

We consider the restriction $H_{D^{\omega},x,L}$ of the operator $H_{D^{\omega}}$ to the cube $\L_L(x)$ with Dirichlet boundary conditions. The MSA is an induction procedure which aims to show the decay of the finite-volume resolvent operator of $H_{D^{\omega}}$ over a sequence of cubes of scales $(L_k)_{k\in\NN}$ that grows to infinity. The main ingredients of the MSA
\begin{itemize}
\item[(ii)] The {\em induction step}: to show that if the finite-volume resolvent decays exponentially between distant points at a certain scale $L_k$ with a good probability, then this is also true at scale $L_{k+1}$. This argument relies on Wegner estimates or similar, and on the geometric resolvent inequality.
 \item[(i)] The {\em initial length scale estimate} (ILSE): the existence of an initial length scale $L_0$ for which the finite-volume resolvent decays exponentially between distant points in space, with good probability. This often relies on the existence of a spectral gap between the first eigenvalue of $H_{D^{\omega},x,L}$ and the spectral infimum of $H_{D^{\omega}}$, and on the Combes-Thomas estimate. The former can be obtained from Lifshitz tails, if this is available.
\end{itemize}
Here, by good probability  we mean a probability that is algebraically close to one in the scale. One we have $(i)$ and $(ii)$, we can apply the MSA which we know to hold in the aperiodic setting \cite{RM12,MRM} and this yields localization at an interval near the spectral infimum.
For a Delone-Anderson operator  localization holds without any conditions on the pattern frequencies of the Delone set. The following was shown in \cite{G,Kl, GMRM, MRM} with different degrees of regularity of the probability distribution $\P_0$.

\begin{thm}[Localization for Delone-Anderson model] Let $D$ be a Delone set and let $H_{D^{\omega}}=-\Delta+V_{D^{\omega}}$ be given in \eqref{hdom}. Then, there exists an energy $E_*$ such that the operator $H_{D^{\omega}}$ exhibits localization in the interval $[0,E_*]$ for almost every $\omega\in \Omega_D$.
\end{thm}
The proof can be obtained in several ways, which differ in the proof of the ILSE and the version of the MSA used. In \cite{G}, this relies on a space averaging and one can allow for arbitrary non-degenerate probability distribution, however the assumption $H_0=-\Delta$ is necessary. In \cite{GMRM} the ILSE relies on averaged bounds on the IDS, of the type Theorem \ref{LT} but that do not require any condition on the Delone set $D$. This requires, however, that the probability measure $\P_0$ is absolutely continuous with bounded and continuous probability density.  The proofs in \cite{Kl,MRM} rely on unique continuation principles and allow for $H_0=-\Delta+V_0$ with $V_0$ a background deterministic potential, see below.
\begin{rem}
\begin{itemize}
\item[i.]We can go further and obtain a lower bound on the length of the interval of localization depending on the parameter $R$ of relative denseness of the Delone set $D$, see \cite[Thm. 3.9]{GMRM}.
    \item[ii.] For an analogous result in the magnetic case \eqref{magnop} without any condition on the Delone set $D$, assuming only H\"older continuity of $\P_0$ see \cite[Thm. 6.1]{RM12}.
    \end{itemize}
\end{rem}

The results on localization stated above rely either on the random variables being regular enough or in the absence of background deterministic potentials. In order to prove localization for the Delone-Anderson model in the case of Bernoulli random variables with a bounded background potential $V_0\neq0$ in Definition \ref{defrandop}, the following quantitative version of the unique continuation principle turns out to be crucial.
\begin{thm}[Quantitative unique continuation]
Let $H_0=-\Delta+V_0$ with $V_0$ a bounded, measurable potential, and $H_{0,L}$ its restriction to the cube $\L_L(x)$ with Dirichlet boundary conditions. Let $E$ be an eigenvalue of $H_{0,L}$ contained in an interval $I$, and let $\psi$ be an eigenfunction associated to $E$. Let $D$ be a Delone set of parameters $(r,R)$ and $0<s<r$. Then, there exists a positive constant $C$ that depends on $s, V, d, R$ and $I$, but does not depend on the scale $L$, such that, uniformly on $x\in \Rd$,
\be \sum_{\gamma\in D\cap \L_L(x)}\norm{\psi}^2_{\L_s(\gamma)}\geq  C\norm{\psi}^2_{\L_L(x)}. \ee
\end{thm}
Here, $\norm{\psi}^2_{\L_s(\gamma)}$ denotes the norm of $\psi$ over the set ${\L_s(\gamma)}$.

This type of estimates, known in the literature for the case $D=\Zd$, were used for the first time to prove localization in the Bernoulli case in the leading-edge work \cite{BK}. Here, the main difficulty is to show that $C$ does not depend on the scale $L$ of the cube. In the case of $D$ aperiodic, this was first obtained in \cite[Thm. 2.1]{RMV}, followed by \cite{Kl}. The fact that this estimate is still valid in the aperiodic case, raises the question of its validity for more general geometric domains $S$, other than cubes $\L_L(x)$. One might also ask if the constant $C$ depends on the geometry of the basic tiles forming the region $S$. Such a result would not only be interesting from the point of view of random operators with underlying Delone structures, but would allow to extend the work done on localization for $N$-particle random Anderson operators in \cite{KlNRM} to the continuous setting. There, the use of cubes defined in a symmetric distance leads to the use of reflected cubes as domains, that form polygons.

We are now ready to formulate our last result on localization for Delone-Anderson models, that will appear in \cite{MRM} and that turns out to be instrumental in the study of purely aperiodic media. Let $D$ and $D'$ be two Delone sets, and consider the Delone-Bernoulli operator given by
\be\label{opber} H_\omega=-\Delta+V_D+V_{D'^\omega}\ee
with
\[ V_D:= \sum_{p\in D}u(\cdot -p), \quad\mbox{and}\quad V_{D'^\omega}= \sum_{\gamma\in D'}\omega_\gamma u(\cdot -\gamma) \]
where the single-site potential $u$ is as in Definition \eqref{defrandop}, with $\P_0$ a Bernoulli distribution satisfying $\P_0(\omega_0=0)=\beta$, for some $\beta\in(0,1)$.
\begin{thm}[Localization for Delone-Bernoulli operator \label{t:locber}] Let $H_\omega$ be the Delone-Bernoulli operator from \eqref{opber}.
\begin{itemize}
\item[i.]  For $d\geq 2$, the operator $H_\omega=-\Delta+V_D+V_{D'^\omega}$ exhibits localization at the bottom of its spectrum, almost surely.
     \item[ii.] If $d=1$, there exists a constant $C>0$ such that if $0<\beta<C$, the the operator $H_\omega=-\Delta+V_D+V_{D'^\omega}$ exhibits localization at the bottom of its spectrum almost surely.
         \end{itemize}
\end{thm}
The proof of this theorem also allows for operators of the form $H_\omega=-\Delta+V_0+V_{D'^\omega}$ with arbitrary non-degenerate probability distribution $\P_0$.
\section{Localization for Delone operators}
In this section, we state some recent results on localization for purely aperiodic media.
We first recall the notion of convergence in the topological space $\DD$ of Delone sets given in \cite[Lemma 3.1]{LS06}.
 \begin{defn}\label{d:conv} A sequence $(D_n)$ of Delone sets converges to $D\in\DD$ in the topology of $\DD$ if and only if there exists, for any $l>0$, an $L>l$ such that the discrete sets $D_n\cap \L_L(0)$ converge to $D\cap \L_L(0)$ with respect to the Hausdorff distance as $n$ tends to infinity.
 \end{defn}
 Given a Delone set of parameters $(r,R)$, we define the Delone operator acting on $\Lp{\Rd}$ by
\be\label{delop}  H_D :=-\Delta+ V_{D}, \ee
with
\begin{equation}
	\label{delop-pot}
 	V_{D} := \sum_{\gamma\in D}u(\cdot-\gamma),
\end{equation}
with $u$ as in Definition \ref{defrandop}.
We denote by $E_D$ the infimum of $\sigma(H_D)$.
The following is a consequence of Theorem \ref{t:locber},
\begin{thm}
Let $D$ be an $(r,R)$-Delone set. Then, there exists an energy $E_*$ and a sequence of Delone sets $(D_n)_{n\in\NN}\subset \mathbb D$ such that $D_n$ converge to $D$ in the sense of Definition \ref{d:conv}, the Delone operators $H_{D_n}$ converge to $H_D$ in the strong resolvent sense, and $[E_D,E_*]$ is contained in the spectrum of the operators $H_{D_n}$, where they exhibit localization.
\end{thm}
The proof consists of associating to $H_D$ the following Delone-Anderson operator with Bernoulli random variables: for $D$, we pick a Delone set $D'$ such that $D\cup D'$ is still a Delone set, and consider the operator $H_\omega=-\Delta+V_D+V_{D'^\omega}$ given by \eqref{opber}. From Theorem \ref{t:locber} we obtain a family of configurations $\hat\Omega\subset \Omega_{D'}$ of full probability measure, such that operators of the form $H_\omega$ with $\omega\in\hat\Omega$ approximate $H_D$ in the sense of Definition \ref{d:conv} and exhibit localization at the bottom of the spectrum. For details, see \cite{MRM}.

This result also allows to show, under certain conditions on the potentials $V_D, V_{D'}$, that configurations $\omega$ associated to localization form a meagre set in the topology of $\Omega_D$, a result that was known for more restrictive random potentials in the work of Simon \cite{Si}. In this sense, our result complements the ones obtained by D. Lenz and P. Stollmann on the singular continuous spectrum of Delone operators in the continuous setting.
\vspace{-0.3cm}
\section*{Acknowledgements}
The author would like to thank F. Germinet and P. M\"uller for their support and a fruitful collaboration. The author would also like to thank the hospitality of the IAM and ECM at the University of Bonn, where part of this note was written.


\begin{thebibliography}{AAAAA}
\bibitem{And58} P.W. Anderson, Absence of Diffusion in Certain Random Lattices, \emph{Phys. Rev.} 109 (1958) 1492--1505.


  \bibitem{BaGr}  M. Baake, U. Grimm, Aperiodic Order. Volume 1: A Mathematical Invitation Encyclopedia of Mathematics and its Applications No. 149, Cambridge University Press, Cambridge.
  \bibitem{BK} J. Bourgain, C. Kenig,  On localization in the continuous Anderson-Bernoulli model in higher dimension,  Invent. Math. 161 (2005) 389--426.

\bibitem{BdMNSS06} A. Boutet de Monvel, S. Naboko, P. Stollmann, G. Stolz,
	Localization near fluctuation boundaries via fractional moments and applications.
    \newblock{J. Anal. Math. 100 (2006) 83--116.}

	

\bibitem{GKuniv} F. Germinet, A. Klein,
\newblock {A comprehensive proof of localization for continuous {Anderson}
  models with singular random potentials}.
\newblock J. Eur. Math. Soc. (JEMS) 15 (2013) 55--143.

\bibitem{GMRM}
F. Germinet, P. M{\"u}ller, C. Rojas-Molina,
\newblock {Ergodicity and Dynamical localization for Delone-Anderson operators}.
\newblock Rev. Math. Phys. 27 (2015) 1550020.

\bibitem{G} Germinet, F.: Recent advances about localization in continuum
random Schr\"odinger operators
with an extension to underlying Delone sets, \emph{Mathematical results in quantum mechanics}, World. Sci. Publ., Hackensack, NJ, 2008, pp. 79--96.


\bibitem{GMP77}
	I. Ya. Gol'dsheid, S. A. Molchanov, L. A. Pastur,
	{A pure point spectrum of the stochastic one-dimensional Schr{\"o}dinger equation},
	\emph{Funkt. Anal. Appl.} 11 (1977) 1--8.
	[Russian original: \emph{Funkts. Anal. Prilozh.} 11 (1977) 1--10].

\bibitem{KLS03a} S. Klassert, D. Lenz, P. Stollmann, Delone dynamical systems: ergodic features and applications.
    \newblock{In {Quasicrystals}, {Structure} and {Physical} {Properties}, Ed. H.R. Trebin. Wiley-VCH Berlin, 2003.}

\bibitem{Kl}   A. Klein, Unique continuation principle for spectral projections of Schr\"odinger operators and optimal Wegner estimates for non-ergodic random Schr\"odinger operators. Comm.Math Phys. 323 (2013) 1229-1246.

\bibitem{KK04} A. Klein, A. Koines, {A general framework for localization of classical waves. II. Random media}, {\em Math. Phys. Anal. Geom.} { 7}  (2004) 151--185.
\bibitem{KlNRM} A. Klein, S.T. Nguyen, C. Rojas-Molina, Characterization of the metal-insulator transport  transition for the two-particle Anderson model, \emph{Ann. Henri Poincaré} 18 (2017) 2327--2365.

\bibitem{LS03} D. Lenz, P. Stollmann, Delone dynamical systems and associated random operators
    \newblock{In Proc. OAMP, Constanta 2001, Eds. Combes, J. M. et al., Theta Foundation, 2003}.


\bibitem{LS06} D. Lenz, P. Stollmann,
 \newblock{Generic sets in spaces of measures and generic singular continuous spectrum for Delone Hamiltonians}
    \newblock{Duke Math. J. 131 (2006) 203--217.}


\bibitem{MR} P. M\"uller, C. Richard, Ergodic properties of randomly coloured point sets, \emph{Canad. J. Math.} 65 (2013) 349--402.
\bibitem{MRM} P. M\"uller, C. Rojas-Molina, {\em in preparation}.





\bibitem{Pas80}
  {L. Pastur},
  {Spectral properties of disordered systems in the one-body approximation},
  \emph{Commun. Math. Phys.} {75} (1980) {179--196}.

  \bibitem{RM12} C. Rojas-Molina, Characterization of the Anderson metal-insulator transition for non ergodic operators and application, \emph{Ann. H. Poincar\'e} 13 (2012) 1575--1611.
\bibitem{RM13}
	C. Rojas-Molina,
	The Anderson model with missing sites,
	\emph{Operators and Matrices} 8 (2014) 287--299.
\bibitem{RMch}
	C. Rojas-Molina,
Random Schrödinger Operators on discrete structures, in Collection S\'eminaires et Congr\'es SMF 32, 147-187, 2018.

  \bibitem{RMV}
Rojas-Molina, C., Veseli\'c, I.: Scale-free unique continuation estimates and applications to random Schr\"odinger operators. Commun. Math. Phys. 320 (2013) 245--274.


  \bibitem{Sch84} D. Shechtman, I. Blech, D. Gratias, J.W. Cahn,
\newblock{Metallic phase with long-range orientational order and no translation symmetry}
\newblock{\em Phys. Rev. Lett.} 53 (1984) 1951--1953.

\bibitem{Si} Simon, B.: Operators with singular continuous spectrum: I. General operators,
\newblock{ Ann. Math. 141, 131--145 (1995)}.


\bibitem{Sto}
	P.~Stollmann,
	\emph{Caught by disorder: Bound states in random media},
	Birkh{\"a}user, Boston, 2001.

 \end{thebibliography}
 \end{document}